\documentclass[aps,prl,showpacs,twocolumn,superscriptaddress]{revtex4}
\usepackage{graphicx}
\usepackage{amssymb}
\usepackage{epsfig}
\usepackage{graphicx}
\usepackage{amsmath}
\usepackage{array,color}

\begin{document}
\title{Spin-orbit Coupling Effects on the
Superfluidity of Fermi Gas in an Optical Lattice}
\author{Q. Sun}
\affiliation{Institute of Physics, Chinese Academy of Sciences,
Beijing 100190, China}
\author{G.-B. Zhu}
\affiliation{Institute of Physics, Chinese Academy of Sciences,
Beijing 100190, China}
\author{W.-M. Liu}
\affiliation{Institute of Physics, Chinese Academy of Sciences,
Beijing 100190, China}
\author{A.-C. Ji}
\email{acji@iphy.ac.cn} \affiliation{Department of Physics, Capital
Normal University, Beijing 100048, China} \affiliation{Institute of
Physics, Chinese Academy of Sciences, Beijing 100190, China}
\date{{\small \today}}

\begin{abstract}
We investigate the superfluidity of attractive Fermi gas in a square
optical lattice with spin-orbit coupling (SOC). We show that the
system displays a variety of new filling-dependent features. At half
filling, a quantum phase transition from a semimetal to a superfluid
is found for large SOC. Close to half filling where the emerging
Dirac cones governs the behaviors of the system, SOC tends to
suppress the BCS superfluidity. Conversely, SOC can significantly
enhance both the pairing gap and condensate fraction and lead to a
new BCS-BEC crossover for small fillings. Moreover, we demonstrate
that the superfluid fraction also exhibits many interesting
phenomena compared with the spin-orbit coupled Fermi gas without
lattice.
\end{abstract}

%67. Quantum fluids and solids
 %67.85 Ultracold gases
   % 67.85.Lm  Degenerate Fermi gases

%03. Quantum mechanics, field theories
  %03.75 matter waves of ultracold  gases
     % 03.75.Ss Degenerate Fermi gases

% 30. ATOMIC AND MOLECULAR PHYSICS
  % 37.10. Atom, molecule, and ion cooling methods
     % 37.10.Jk Atoms in optical lattices

%70. CONDENSED MATTER: ELECTRONIC STRUCTURE, ELECTRICAL, MAGNETIC
    %71. Electronic structure of bulk materials
       % 71.10.Fd Lattice fermion models (Hubbard model, etc.)

%05. Statistical physics
  % 05.30.Fk Fermion systems and electron gas

%71.70.Ej Spin ¨Corbit coupling, Zeeman and Stark splitting
\maketitle

The spin-orbit coupling (SOC) plays a central role in the
investigation of novel topological states in solid state physics
\cite{Hasan,Qi}.  This has stimulated tremendous interests in
creating artificial non-Abelian gauge fields in ultracold atom
systems \cite{Dalibard}. The successful realization of SOC in both
Bose-Einstein condensate (BEC) \cite{Lin1,Lin2} and Fermi gas
\cite{Wang2,Cheuk} opens up a new avenue towards studying the rich
physics of spin-orbit (SO) coupled ultracold atoms
\cite{Stanescu,Wang,Ho,Sinha,Iskin,Seo}. One of the important
advances is that SOC was shown to have fundamental effects on the
superfluidity of 3D \cite{Vyasanakere,Hu, Yu, Gong,Han,Zhou} and 2D
\cite{Zhou,He} continuous Fermi gases.

On the other hand, the attractive Fermi gas subjected to an optical
lattice \cite{Esslinger,Bloch} has made it possible to simulate the
negative-$U$ Hubbard model, a basic model for the superconductivity
of many solid state materials \cite{Micnas}. In particular, the
on-site attractions can induce deep bound states, which cause the
conventional BCS-BEC crossover. Recently, SOC has been combined to
optical lattices for repulsive ultracold gases and predicted to lead
to many interesting phenomena \cite{Radic,Cole,Grab}. Nevertheless,
the superfluidity of SO coupled attractive Fermi gas in an optical
lattice remains a new frontier to be explored.

In this Letter, we study the Fermi gas subjected to a square optical
lattice with SOC. Such a system can be described by a generalized
negative-$U$ Hubbard model. We show that, the combination of SOC and
lattice can give rise to various new features that depend on the
fillings. Remarkably, there develops a quantum phase transition
(QPT) from a semimetal to a superfluid for large SOC at half
filling,  with the critical interaction $U_c/t\simeq3.11$ ($t$ is
the hopping amplitude). For close to half filling, we show that the
emerging Dirac cones governs the behaviors of the system, which
tends to suppress the BCS superfluidity. By contrary, SOC can
significantly enhance both the pairing gap and condensate fraction
and lead to a new BCS-BEC crossover for small fillings. Compared
with the SO coupled Fermi gas without lattice, such opposite
filling-dependent behavior of SOC is rather unique as it can only be
induced in the lattice system. Furthermore, we investigate the
superfluid fraction, which also exhibits many unusual
characteristics in contrast to the continuous Fermi gas.

\begin{figure}[h]
\includegraphics[width=0.43\textwidth]{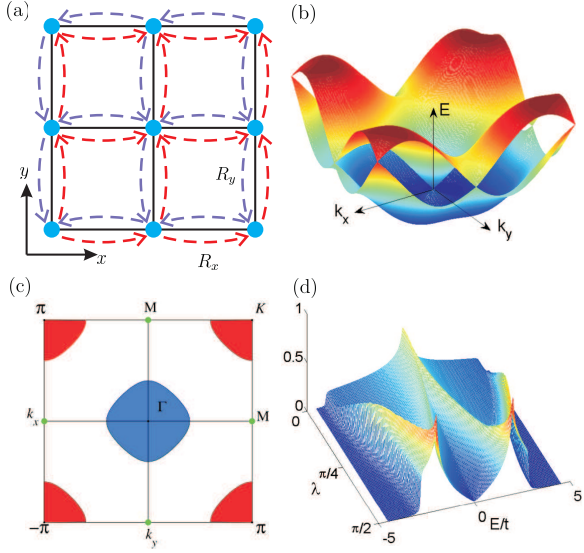}
\caption{(color online) (a) SO coupled square optical lattice, where
$R_{x(y)}$ denotes the non-Abelian hopping matrice along $x(y)$
direction. (b) Energy bands of non-interacting fermions with
$\lambda=3\pi/10$ for illustration. (c) Zero energy Fermi surfaces
at half filling, where the filled particle (blue) and hole (red)
Fermi pockets represent the up/down Rashba  band respectively. The
green dots denote two zero energy Fermi points at $M$. (d) Density
of states $\rho(E)$ over the regime $\lambda\in[0,\pi/2]$.}
\label{fig1}
\end{figure}

We consider a system of two-component Fermi gas moving in an optical
square lattice. In the tight binding approximation, the Hamiltonian
reads
\begin{eqnarray}
H&=&-t\sum_{<ij>}\sum_{\sigma\sigma^\prime}
(c^\dag_{i\sigma}R_{ij}c_{j\sigma^\prime}+{\rm H.c.})\nonumber\\
&&-U\sum_{i} n_{i\uparrow}n_{i\downarrow}-\mu\sum_{i} n_{i},
\label{Hamiltonian}
\end{eqnarray}
where $t$ is the overall hopping amplitude and $c^\dag_{i\sigma}$ is
the creation operator for spin-up (down) fermion
$\sigma=\uparrow,\downarrow$ at site $i$. The nearest sites
tunneling matrices $R_{ij}= e^{i\vec{A}\cdot (\vec{r}_j-\vec{r}_i)}$
with $\vec{A}=\lambda\left(\sigma _{x},\sigma _{y}\right)$ the
non-Abelian gauge-field \cite{Radic,Cole,Grab,Goldman}, $\lambda$ is
the strength of Rashba SOC \cite{Note1} (see Fig. \ref{fig1}(a)).
Here, the diagonal term of $R_{ij}$ denotes the spin-conserved
hopping, while the non-diagonal term can be realized by the Raman
laser assisted spin-flipped tunneling \cite{Osterloh}. $U$ is the
on-site attraction strength which can be tuned by Feshbach
resonances and $\mu$ is the chemical potential. $n=\langle
n_{i\uparrow}+n_{i\downarrow}\rangle$ is the filling factor.

Fig. \ref{fig1}(b) shows the  band structure of non-interacting
fermions, where SOC lifts the spin degeneracy and gives rise to two
split Rashba bands. Remarkably, the two bands intersect linearly at
$\Gamma=(0,0)$, $M=(\pi,0),(0,\pi)$ and $K=(\pi,\pi)$.  The zero
energy Fermi surfaces at half filling is shown in Fig.
\ref{fig1}(c), where we have a particle (hole) Fermi-pocket around
$\Gamma$ ($K$) which is associated to the up (down) Rashba band
respectively. Note that, there always exist two zero energy Fermi
points at $M$ for any $\lambda\neq0$. Specifically, when
$\lambda=\pi/2$, both the particle (hole) Fermi-pockets shrink to
Fermi points at zero energy, and there develops a semimetal with
four Dirac cones at $\Gamma,K$ and $M$. Fig. \ref{fig1}(d) shows the
density of states (DOS) $\rho(E)$ of single-particle excitation over
the regime $\lambda\in[0,\pi/2]$, we see that when $\lambda=\pi/2$,
$\rho(E)\propto |E|$ which vanishes linearly around zero energy.

We start by writing the partition function in the imaginary-time
path integral $\mathcal{Z}=\int\mathcal{D}[\bar{\psi},\psi]
e^{-S[\bar{\psi},\psi]}$, where
$S[\bar{\psi},\psi]=\int^\beta_0d\tau[\sum_\sigma\bar{\psi}_\sigma
\partial_\tau\psi_\sigma+H(\bar{\psi},\psi)]$
with $\psi=(\psi_\uparrow,\psi_\downarrow)^T$ representing the
Grassmann field variables. Then, by decoupling the attractive term
of Eq. (\ref{Hamiltonian}) in normal and anomalous channels through
a pairing field
$\Delta_i(\tau)=U\psi_{i\downarrow}(\tau)\psi_{i\uparrow}(\tau)$ and
introducing $(\psi_{\mathbf{k}\uparrow},\psi_{\mathbf{k}\downarrow},
\bar{\psi}_{-\mathbf{k}\uparrow},\bar{\psi}_{-\mathbf{k}\downarrow})^T$,
we obtain the effective action after integrating out the fermionic
field $S_{\rm{eff}}=\sum_i\int^\beta_0
d\tau\frac{|\Delta_i(\tau)|^2}{U}-\frac{1}{2}\rm{Tr}\ln
\mathcal{G}^{-1}+\beta\sum_\mathbf{k}\varepsilon_\mathbf{k}$. Here
we have ignored the constant term $NUn^2/4$ ($N$ is the number of
lattice sites) and the inverse Green function is given by
\begin{eqnarray}
\mathcal{G}^{-1}=\left(
\begin{array}{cc}
\partial_\tau\>+\>\varepsilon_\mathbf{k}
\>+\>\lambda_\mathbf{k}\>\>\>\> & -i\sigma_y\Delta_i(\tau) \\
i\sigma_y\bar{\Delta}_i(\tau)\>\>\>\> & \partial_\tau
\>-\>\varepsilon_\mathbf{k}\>+\>\bar{\lambda}_\mathbf{k}\\
\end{array}
\right),
\end{eqnarray}
with $\varepsilon_\mathbf{k}=-2t\cos\!\lambda(\cos k_x+\cos
k_y)-\bar{\mu}$ and $\lambda_\mathbf{k}=-2t\sin\!\lambda(\sin
k_x\sigma_x+\sin k_y\sigma_y)$, where $\bar{\mu}=\mu+Un/2$ is the
scaled chemical potential. Furthermore, we set
$\Delta_i(\tau)=\Delta+\delta\Delta$ and write
$\mathcal{G}^{-1}=\mathrm{G}^{-1}+\Sigma$ with
$\mathrm{G}^{-1}=\mathcal{G}^{-1}|_{\Delta_i(\tau)=\Delta}$. Then,
the effective action can be expanded to the second order of
fluctuation $\Sigma$ as  $S_{\rm{eff}}\simeq S_0+\Delta S$ with
$S_0=\frac{\beta
N}{U}\sum|\Delta|^2+\frac{1}{2}\sum_{\mathbf{k},\nu=\pm}[\frac{\beta}{2}
(\varepsilon_\mathbf{k}-E_{\mathbf{k},\nu})-\ln(1+e^{-\beta
E_{\mathbf{k},\nu}})]$, and $\Delta
S\equiv\sum_q\Gamma^{-1}(q)\delta\bar{\Delta}(-q)\delta\Delta(q)
=\frac{N}{U}\sum_q\delta\bar{\Delta}(-q)\delta\Delta(q)
+\frac{1}{4}{\rm{Tr}}[\mathrm{G}(k)\Sigma(-q)\mathrm{G}(k-q)
\Sigma(q)]$. Here $k=(\mathbf{k},iw_n)$, $q=(\mathbf{q},i\nu_n)$,
and $E_{\mathbf{k},\pm}=\sqrt{\xi_{\mathbf{k},\pm}^2+\Delta^2}$ with
$\xi_{\mathbf{k},\pm}=\varepsilon_\mathbf{k}\pm2t
\sin\!\lambda\>\mathcal{K}$ being the two Rashba branches,
$\mathcal{K}\equiv\sqrt{\sin^2 k_x+\sin^2 k_y}$.  At the mean-field
level, the many-body ground state of the system can be derived by
minimizing $S_0/(N\beta)$ with respect to $\Delta$ and $\mu$, and we
have the following gap and Fermi density equations
\begin{eqnarray}
\frac{1}{U}&=&\frac{1}{N}\sum_{\mathbf{k},\nu=\pm}
\frac{1}{4E_{\mathbf{k},\nu}}\tanh(\frac{\beta
E_{\mathbf{k},\nu}}{2}),\nonumber\\
n&=&1-\frac{1}{N}\sum_{\mathbf{k},\nu=\pm}\frac{\varepsilon_\mathbf{k}}
{2E_{\mathbf{k},\nu}}\tanh(\frac{\beta E_{\mathbf{k},\nu}}{2}).
\end{eqnarray}

Before proceeding, it's useful to consider the large attraction
limit with $U/t\gg1$. In this case, the standard degenerate
perturbation theory can be applied for Eq. (\ref{Hamiltonian})
through the canonical transformation $c_{i\uparrow}\rightarrow
c_{i\uparrow}$ and $c_{i\downarrow}\rightarrow(-1)^{i_x+i_y}
c_{i\downarrow}^\dagger$ \cite{Robaszkiewicz}. For any band filling,
we can derive an effective spin model
$H_{\rm{spin}}=J\sum_{<ij>}\mathbf{S}_i\cdot\mathbf{S}_j -2\bar
{\mu}\sum_i S^z_i$ in the presence of {\it arbitrary SOC}. Here
$J=4t^2/U$ and the pairing field operator becomes the transverse
magnetic operator. For $\bar {\mu}\neq0$, the antiferromagnetic
order in $XY$ plane is equivalent to the pairing order of  Eq.
(\ref{Hamiltonian}). Therefore, we conclude that SOC does not have
any influence on the superfluidity of the system in the large $U$
limit, where all the fermionic atoms form tightly bound molecules
and give rise to a Kosterlitz-Thouless transition of BEC
\cite{Scalettar}. In the following, we shall focus on the more
interested weak and intermediate attraction regions.

\begin{figure}[h]
\includegraphics[width=0.49\textwidth]{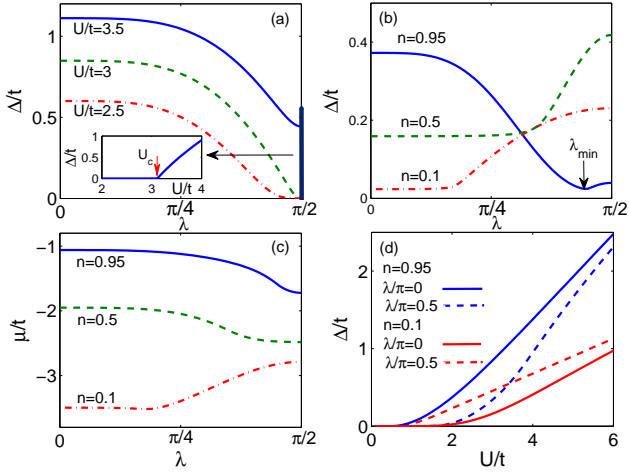}
\caption{(color online) (a) Plot of $\Delta$  versus $\lambda$ at
half filling for different $U/t$. Inset shows a semimetal-superfluid
QPT with $U_c/t\simeq3.11$. (b) Plot of $\Delta$ and (c) $\mu$
versus $\lambda$ with $U/t=2$ for different fillings.  (d) Pairing
gap as a function of $U/t$ for two typical fillings with $n=0.1$
(red) and $0.95$ (blue). Solid and dashed lines represent
$\lambda=0$ and $\pi/2$ respectively.}\label{fig2}
\end{figure}

The pairing gaps at zero temperature are illustrated in  Fig.
\ref{fig2}. First for half filling ($n=1$), Fig. \ref{fig2} (a)
shows that the BCS gap decreases monotonically with respect to
$\lambda$. This could be understood that, the Fermi pockets around
$\Gamma$ and $K$ (see Fig. \ref{fig1}(c)) tend to form the Fermi
points at $E_F=0$ by increasing SOC, which causes a suppression of
DOS at zero energy (see Fig. \ref{fig1}(d)). Specifically, when
$\lambda=\pi/2$ the system becomes a semimetal, which is expected to
be stable towards small attractions. On the other hand, when
$U/t\gg1$ the system should support a superfluid state of bound
molecules as indicated by the effective spin model $H_{\rm{spin}}$
\cite{Note2}. Hence, there must undergo a significant QPT from a
semimetal to a superfluid by increasing attractions, see the thick
vertical line of Fig. \ref{fig2} (a). In the inset, we show that the
critical value $U_c/t\simeq3.11$, above which a finite gap develops.

\begin{figure}[h]
\includegraphics[width=0.44\textwidth]{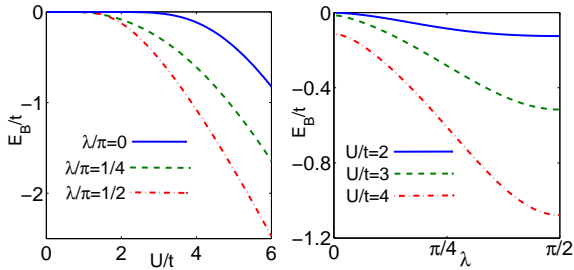}
\caption{(color online) Binding energy $E_B/t$ as a function of
$U/t$ (left panel) and SOC strength $\lambda$ (right panel).
}\label{fig3}
\end{figure}

Such scenario would be affected remarkably by dopings as shown in
Fig. \ref{fig2} (b). Without loss of generality, We focus on the
hole doping case due to the particle-hole symmetry of the system.
First for small dopings, similar to that of half filling, the
superfluidity is governed by emerging Dirac cones at zero energy and
$\Delta$ is suppressed by increasing $\lambda$, see $n=0.95$.
However, when close to $\lambda=\pi/2$, the doping would make the
QPT at half filling unstable and opens a gap. This produces a
nonmonotonic behavior of $\Delta$ with a minimum at $\lambda_{\rm
min}$. While for large dopings, the situation is entirely changed,
where the influence of Dirac cones would diminish and SOC induces a
new BCS-BEC crossover with $\Delta$ being significantly enhanced,
see the dash dotted line of $n=0.1$. This can be understood by
solving the two-body problem of Eq. (\ref{Hamiltonian}), which is
determined by
$\Gamma^{-1}(i\nu_n\rightarrow\omega+i0^+,\mathbf{q}=0)=0$ as
$\omega+2{\bar\mu}=-E_B$ and we arrive at
\begin{eqnarray}
\frac{1}{U}=\frac{1}{2N}\sum_{\mathbf{k},\nu=\pm}
\frac{1}{2(\xi_{\mathbf{k},\nu}-E_0)-E_B}, \label{Binding}
\end{eqnarray}
where $E_0$ denotes the lowest energy of $\xi_{\mathbf{k},-}$
branch. In the absence of SOC ($\lambda=0$), the  binding energy
$|E_B|/t\sim0$ in the weak attraction region $U/zt<1$ ($z=4$ is the
number of the nearest neighbor) and becomes very large for
$U/zt\gg1$, evolving from loosely local pairs (BCS) to tightly bound
molecules (BEC) \cite{Micnas}. However, when SOC is added to the
lattice, $|E_B|/t$ will be significantly enhanced (left panel of
Fig. \ref{fig3}) due to the increasing of DOS around $E_0$ and
effective shrinking
of the bandwidth (see Fig. \ref{fig1}(d))
\cite{Note3}. In particular, the right panel shows a remarkable grow
of $|E_B|/t$ from nearly zero in the weak attraction regions, which
signifies the formation of SOC induced bound states, see $U/t=2,3$
for example.

In general, such filling-dependent effects arise from the unique
features of combination of SOC and lattice. The opposite behaviors
of $\Delta$ versus SOC upon dopings indicate that the system evolves
from the Dirac cone dominated physics near half filling to the SOC
induced BCS-BEC crossover at small fillings. In this respect, the
gap behavior at small fillings is reminiscent of SO enhanced pairing
in the unitary Fermi gas \cite{Vyasanakere,Hu,Yu, Gong,Han,Zhou,He}.
However, we note that the lattice also plays nontrivial roles even
in this limit. In Fig.\ref{fig2} (c), we see that $\mu$ increases
with SOC at $n=0.1$, which differs from the familiar results in the
continuous system where $\mu$ is decreased by SOC.

The opposite roles of SOC in lattice can be clearly seen in Fig.
\ref{fig2} (d), where we plot $\Delta$  as a function of  $U/t$ for
two typical fillings. We take $\lambda=\pi/2$ for illustration and
show that the strong SOC can remarkably  enhance ($n=0.1$) or
suppress ($n=0.95$) the pairing gaps of the conventional BCS-BEC
crossover of negative-$U$ Hubbard model, especially in the weak and
intermediate attraction regions. While in the large attraction
limit, $\Delta$ versus $U/t$ approaches the $\lambda=0$ results,
which means the SOC effects diminish according to the effective
theory of $H_{\rm{spin}}$.

To get more insight of the unusual properties of this system, we now
explore the condensate and superfluid fractions. First, the
condensate density $n_c=\frac{1}{N}\sum_{{\bf
k},\sigma,\sigma^\prime}|\langle\psi_{\mathbf{k}\sigma}
\psi_{\mathbf{-k}\sigma^\prime}\rangle|^2$ \cite{Gorkov}, where the
singlet and induced triplet pairing fields
$\langle\psi_{\mathbf{k}\uparrow}
\psi_{\mathbf{-k}\uparrow}\rangle=-\frac{\Delta}{4}
e^{-i\theta_\mathbf{k}}\sum_\nu\frac{\nu}{E_{\mathbf{k},\nu}}$ and
$\langle\psi_{\mathbf{k}\uparrow}
\psi_{\mathbf{-k}\downarrow}\rangle
=-\frac{\Delta}{4}\sum_\nu1/E_{\mathbf{k},\nu}$ with
$\theta_\mathbf{k}=\arg(\sin k_x+i\sin k_y)$. While for the
superfluid density, we impose a phase twist on order parameter
$\Delta\rightarrow\Delta e^{i\nabla\theta\cdot \vec{r}_j}$ by a
local unitary transformation $\psi_j\rightarrow
\psi_je^{i\theta(\vec{r}_j)}$. Then, the inverse Green function can
be written as $\mathbf{G}^{-1}[\Delta,\nabla\theta]=\mathrm{G}^{-1}
[\Delta]+\Sigma[\nabla\theta]$. After lengthy but straightforward
calculations, we derive a classical phase variation model
$\widetilde{H}=\frac{1}{2}J\int d^2{\bf
r}[(\partial_x\theta)^2$$+(\partial_y\theta)^2]$ with $J$ the phase
stiffness. Therefore, the superfluid density can be defined as
$\rho_s=\frac{J}{2tN}$, which reads
\begin{eqnarray}
&&\rho_s =\nonumber\\
&&\frac{\cos\!\lambda}{N}\!\sum_\mathbf{k}\cos k_xn_\mathbf{k}\!+\!
\frac{\sin\!\lambda}{N}\!\sum_{\mathbf{k},\nu}
\frac{\nu\xi_{\mathbf{k},\nu}}{2E_{\mathbf{k},\nu}}
\frac{\sin^2k_x}{\mathcal{K}}\! \tanh(\frac{\beta
E_{\mathbf{k},\nu}}{2})\nonumber\\
&&+\frac{2t}{N} \sum_{\mathbf{k},\nu}f^\prime(E_{\mathbf{k},\nu})
\sin^2k_x\left(\cos\lambda+\nu\frac{\sin\lambda\cos
k_x}{\mathcal{K}}\right)^2\nonumber\\
&&- \frac{\sin\lambda}{N}\sum_{\mathbf{k},\nu}
\nu\frac{\varepsilon^2_\mathbf{k} +\nu 2t
\sin\!\lambda\>\mathcal{K}\varepsilon_\mathbf{k} +\Delta^2}{2
\varepsilon_\mathbf{k}E_{\mathbf{k},\nu}}
\frac{\sin^2k_y\cos^2k_x}{\mathcal{K}^3}
\nonumber\\
&&\>\>\>\>\times\tanh(\frac{\beta E_{\mathbf{k},\nu}}{2}).
\label{superfluidity}
\end{eqnarray}
Here $n_\mathbf{k}=1-\sum_{\nu=\pm}\frac{\varepsilon_\mathbf{k}}
{2E_{\mathbf{k},\nu}}\tanh(\frac{\beta E_{\mathbf{k},\nu}}{2})$ and
the third term vanishes at $T=0$.
\begin{figure}[h]
\includegraphics[width=0.5\textwidth]{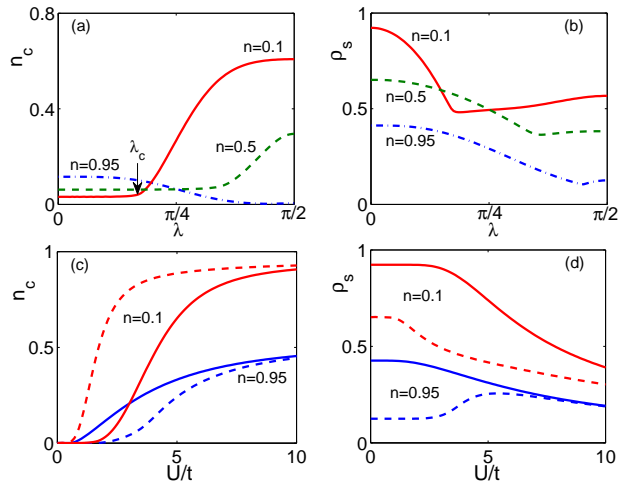}
\caption{(color online) (a) Condensate density $n_c$ and (b)
superfluid density $\rho_s$ (both divided by $n$) at $T=0$ as a
function of SOC for different fillings, where we take $U/t=2$. (c)
Plot of $n_c$ and (d) $\rho_s$ versus $U/t$ for two typical fillings
with $n=0.1$ (red) and $0.95$ (blue). Solid and dashed lines
represent $\lambda=0$ and $\pi/2$ respectively.}\label{fig4}
\end{figure}
Not that, although the first and
fourth terms bear similarities with the continuous system
\cite{Zhou,He}, they may become negative in the lattice.
Fortunately, the new second term which exists only in the SO coupled
lattice system, can stabilize the superfluidity and gives rise to
many intriguing features.

Fig. \ref{fig4} (a) shows the condensate fraction $n_c$ versus SOC
with the evolution of fillings. For $n=0.1$, we see that $n_c$
increase dramatically above a characteristic value $\lambda_c$,
which agrees with the formation of the SOC induced bound states. The
characteristic value $\lambda_c$ grows with increasing fillings, and
until $n\simeq0.7$, $n_c$ begins to decrease with respect to
$\lambda$. This indicates that the BCS superfluidity would be
suppressed at large fillings, see $n=0.95$ for example. On the
contrary, the superfluid fraction $\rho_s$ is always suppressed by
SOC and generally decreased with increasing $n$, as shown in Fig.
\ref{fig4} (b). Significantly, there exhibits a characteristic
minimum of $\lambda$, which moves rightwards when $n$ is increased.

In Fig. \ref{fig4} (c) and (d), we plot $n_c$ and $\rho_s$ with
respect to $U/t$ for two typical fillings. Qualitatively different
from the conventional BCS-BEC crossover, the condensate fraction is
remarkably enhanced ($n=0.1$) or suppressed ($n=0.95$) by the strong
SOC. Conversely, the superfluid fraction is destroyed by both
increasing $U/t$ and $\lambda$. Intersetingly, $\rho_s$ exhibits
quite a nontrivial behavior at large fillings. The presence of
strong SOC can dramatically suppress $\rho_s$ for the weak
attractions and causes a maximum of $\rho_s$ located at the
intermediate crossover region, see $n=0.95$ for example. While in
the large attraction limit, both $n_c$ and $\rho_s$ will approach
the results without SOC.

In summary, we have shown that the SO coupled Fermi gas in an
optical lattice displays various new filling-dependent features. At
half filling, we find a QPT from a semimetal to a superfluid for
large SOC. While upon dopings, the system evolves from the Dirac
cone dominated physics near half filling to the SOC induced BCS-BEC
crossover at small fillings. Moreover, we show that all the pairing
gap, condensate and superfluid fractions exhibit many interesting
physics, which differ qualitatively from the SO coupled Fermi gas
without lattice and the conventional negative-$U$ Hubbard model
without SOC. We hope that this work will trigger new exciting
interests to the SO coupled optical lattice physics, and may be
useful for the study of superconductivity of future solid state
materials with SOC.

\begin{acknowledgments}
We acknowledge Hui Zhai, G. Juzeli\={u}nas, and X. F. Zhang for
helpful discussions. We are grateful to Hui Zhai for reading the
manuscript. This work is supported by NCET, NSFC under grants Nos.
11074175,  10934010, NSFB under grants No.1092009, NKBRSFC under
grants Nos. 2011CB921502, 2012CB821305, and NSFC-RGC under grants
No. 11061160490.
\end{acknowledgments}

\end{document}